\documentclass[newstyle,guide]{rmaa}
\usepackage{rmaacite,epsfig}
  
\def\mp{$M_{\odot}$/pc$^2$}
      
\title{The Milky Way and the evolution of disk galaxies}
\author{N. Prantzos\altaffilmark{1}
  \affil{Institut d'Astrophysique de Paris} 
  }
\altaffiltext{1}{Email: \texttt{prantzos@iap.fr}}
\fulladdresses{
\item 
}
    
\shortauthor{Prantzos} \shorttitle{\RMAA\ Author Guide}

\ReceivedDate{1999 May 26} 
\AcceptedDate{2000 April 13} 
\SetYear{2000}
\resumen{}

\abstract{We present a phenomenological approach to the study of
disk galaxy evolution, based
on i) a detailed modelling of the Milky Way (used as a prototype disk galaxy)
and ii) an extension of the model to other disks through some simple scaling 
relations, obtained in the framework of Cold Dark matter models. The main conclusion
is that, on average,  massive disks have formed the bulk of their stars earlier
than their lower mass counterparts. It is not yet clear why the ``star formation
hierarchy'' has been apparently opposite to the ``dark matter assembly'' hierarchy.
}

\begin{document}
\maketitle

\section{Introduction}
\label{sec:intro}

Accordingly to the currently popular scenario for galaxy formation (pioneered
by White and Rees 1978) the dark haloes of galaxies form hierarchically by
the gravitational clustering of non-dissipative dark matter, 
while the luminous parts form through a combination of gravitational 
clustering and dissipative collapse (which may be affected by feedback).
The major uncertainty in this scenario concerns the baryonic component, since 
the physics of star formation and feedback are very poorly understood at 
present, thus requiring a parametric approach to the problem.
Semi-analytic models of galaxy formation help to explore large regions of the
relevant parameter space and have produced quite encouraging results
(e.g. Cole et al. 2000); still, they have not yet managed to reproduce 
succesfully some key observed properties of disks (van den Bosch 2002) or
ellipticals (Thomas et al. 2002).

Simple (i.e. not dynamical) models of (chemical and/or photometric)
galaxy evolution use, in general, fewer free parameters than semi-analytical
models. If the number of observables that are successfully reproduced is 
(much) larger than the number of free parameters, it is reasonable to assume 
that the galaxian histories produced by such models may indeed 
match the real ones.  In that spirit, we have developed a simple approach to disk 
galaxy evolution, based on i) a detailed model of the Milky Way  (used as
a prototype, Sec. 2) and ii) an extension to other disks trough some simple scaling
relations (Sec. 3). The main conclusion (Sec. 4) is that massive disks have formed
the bulk of their stars earlier than low mass ones. Whether this ``star formation
hierarcy'' is compatible with the ``dark matter assembly'' hierarchy remains to be
demonstrated.

\section{A simple model for the Milky Way disk}
\label{sec:MilkyWay}

Several simple models have been developed in the 
past few years on the chemical 
evolution of the Milky Way disk (e.g. Prantzos and Aubert 1995,
Chiappini et al. 1997, 
Boissier and Prantzos 1999, Chang et al. 1999, Portinari and Chiosi 2000). 
Although they may differ considerably
in their ingredients (radial dependence of Star formation rate and/or infall
rate, inclusion or not of radial inflows, etc.) these models converge in the
following points (see e.g. discussion in Tosi 2000): 
i) infall (of primordial or low metallicity gas) on a long timescale ($\sim$7
Gyr) is required locally, in order to reproduce the observed  G-dwarf
metallicity distribution in the solar neighborhood; ii) strong radial 
dependence of the Star formation rate and/or the infall rate is
needed in order to  reproduce the observed radial profiles of gas, SFR 
and metal abundances [Note: in recent cosmological hydro-simulations of 
galaxy formation Sommer-Larsen et al. (2002) find that gas accretion in 
regions 8 kpc distant from the centers of Milky Way type spirals takes place
on timescales of $\sim$6 Gyr, in reasonable agreement with the requirements of
simple models].

Fig. 1 displays the results of  a simple model, 
(from Boissier and Prantzos 1999)
which also reproduces all the main observables in the solar neighborhood.
The adopted SFR as a function of radius $R$ is of the form
\begin{equation}
\Psi(R) \ = \ A \ V(R) \ R^{-1} \ \Sigma_G^{1.5}(R)
\end{equation}
where $V(R)$ is the rotational velocity at $R$
and  $\Sigma_G$ the gas surface density; this SFR
is based on the assumption that large scale star formation in spirals
is induced by the passage of the spiral waves (Wyse and Silk 1989). 

\begin{figure}[!t]
\begin{center}
\psfig{figure=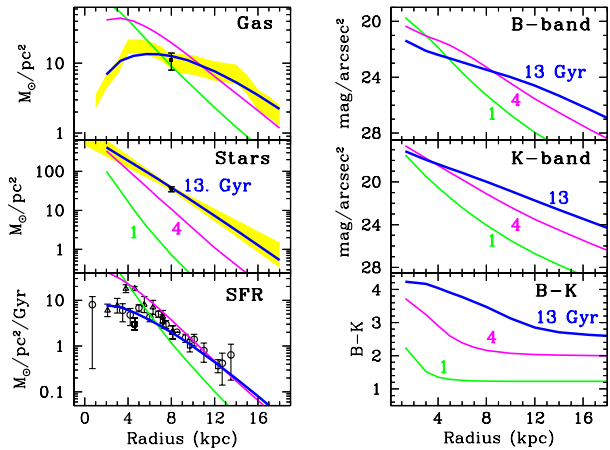,angle=-90,width=0.5\textwidth}
\caption{Chemical ({\it left}) and photometric ({\it right}) 
evolution of the Milky Way
disk, according to the model of Boissier and Prantzos (1999). In all panels
the solid curves correspond to model results at galactic ages of 1, 4 and 13
Gyr, respectively;  the latter ({\it heavy curves}) 
are compared to observations of the present
day disk (in the left panels: {\it shaded} regions for the gaseous and stellar
profiles and {\it data points} for the Star Formation Rate). The model leads
naturally to different scalelengths for the B-band (4 kpc) and the K-band
(2.6 kpc), in agreement with observations.
}
\end{center}
\end{figure}

\begin{figure}
\begin{center}
\psfig{figure=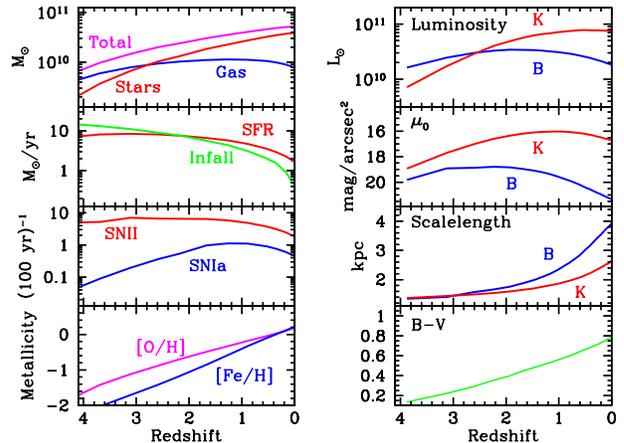,angle=-90,width=0.5\textwidth}
\caption{Results of the Milky Way disk model of Boissier and Prantzos (1999)
plotted as a function of the redshift (for a Universe with $\Omega_M$=0.30,
$\Omega_{\Lambda}$=0.70 and H$_0$=65 km/s/Mpc). The model suggests that
such disks are fainter in the K-band in the past, but slightly brighter in the
B-band at intermediate redshifts; they are also more compact in the past,
especially in the B-band.
}
\end{center}
\end{figure}

The model reproduces satisfactorily the observed chemical and photometric 
radial profiles of the MW disk (Fig. 1) as well as those of oxygen 
and other metal
abundances (Hou et al. 2000); in particular, the model predicts  a flatenning
of the abundance gradients with time (due to the assumed
inside-out formation of the 
disk), in agreement with the recent observational results of Maciel and 
Da Costa(2002),
based on O abundances of planetary nebulae of various age classes 
[Note: taking into account the 
uncertainties in evaluating ages of planetary nebulae, the importance of that
agreement should not be overestimated; it is, however, encouraging].

Fig. 2 displays the evolution of several observables of a Milky Way disk
as a function of redshift in a ``standard'' cosmology, according to the
model. It appears that such disks evolved with quasi-constant amount
of gas and were in the past substantially more compact
in the B-band; they also were  slightly brighter in the B-band 
(at least at redshifts $z\sim$
1-2) and always  fainter in the K-band.

Although the model reproduces successfully the present day profiles of the 
Milky Way disk, its ``predictions'' for the past history of such disks cannnot
be considered very robust, since a degenerate solution (several different
histories leading to the same final state) cannot be excluded. 
Still, it is tempting to explore the consequences of that model, after  
extending it to other disk galaxies.


\section{Scaling relations for galactic disks}

Assuming  that the MW is a typical spiral, one may try to extend the model 
to the case of other disk galaxies by means of simple scaling relations
(in an analoguous way, the Sun is a typical star used to ``calibrate'' 
models of stellar evolution).
The scaling relations derived by Mo et al. (1998) in the framework 
of Cold Dark Matter models of galaxy formation can be used 
(Boissier and Prantzos 2000, BP2000) to describe disks
as two-parameter family (scale length $R_d$ and central surface brightness 
$\Sigma_0$), in terms of the corresponding MW parameters 
(represented by subscript G below):

\begin{equation}
\frac{R_d}{R_{dG}}  \  = \  \frac{\lambda}{\lambda_G} \ \frac{V_C}{V_{CG}}
\end{equation}
and
\begin{equation}
\frac{\Sigma_0}{\Sigma_{0G}}  \  = \  \frac{m_d}{m_{dG}} \
 \left(\frac{\lambda}{\lambda_G}\right)^{-2}
 \ \frac{V_C}{V_{CG}} 
\end{equation}
The two fundamental parameters entering these relations are the circular
velocity $V_C$ (a measure of the disk's mass) and the spin 
parameter $\lambda$ (a measure of the disk's angular momentum
distribution); large $V_C$ values correspond to massive
disks and large $\lambda$ values to extended ones. The baryonic fraction $m_d$
of the disks is taken to be 0.05 in our calculations.
Finally, for the MW parameters
we have: $R_{dG}$=2.6 kpc, $\Sigma_{0G}$=1150 \mp, $V_{CG}$=220 
km/s.

Note that these scaling relations apply to gaseous disks, assumed to be formed
at the redshift of the corresponding dark halo formation; this formalism is
adopted in semi-analytic models of galaxy evolution, where the baryon
accretion history (the ``infall'' history, in our terminology)
is determined essentially by the dark matter accretion history. A Schmidt
type law for star formation (with an efficiency proportional to local 
dynamical timescale) is usually assumed in such models, along 
with some prescription for ``feedback'' (a term describing 
the way gas is affected by the energy input from stars); the interplay between
the three effects (accretion, star formation, feedback) determines
the effective star formation efficiency in semi-analytic models. 


\begin{figure}[!t]
\begin{center}
\psfig{figure=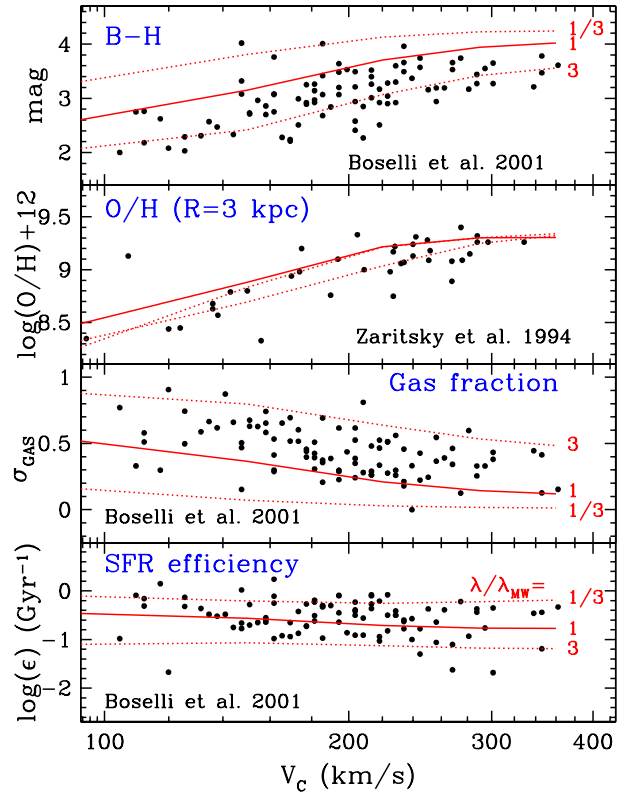,width=0.5\textwidth}
\caption{From top to bottom: Colour, metallicity at 3 kpc from the center,
gas fraction and star formation efficiency (defined as star formation rate
divided by the total gas amount) as a function of rotational velocity for
galactic disks. In general, more massive (rapidly rotating) disks are
redder, more metal-rich and gas poor than  less massive ones; however, the
star formation efficiency does not seem to depend on the disk's mass.
Put together, these observational data suggest that the more massive disks
are chemically and photometrically older than less massive ones. The curves
corrspond to model results of Boissier et al. (2001) for three values of
the spin parameter $\lambda$: 1/3, 1 and 3 times the one of the Milky Way
(the latter corresponds to the solid curves in all panels).
}
\end{center}
\end{figure}

A different approach is adopted in BP2000:
the star formation law and efficiency A (Eq. 1)
are kept the same as in the MW model
(with rotational velocity curves properly calculated),
while the infall rate is adjusted in order to reproduce main properties
of present-day spirals.
This approach (i.e. using present-day properties of galaxies to infer their
histories) has been characterised as ``backwards'' approach to galaxy evolution
(e.g. Fereras and Silk 2001); our use of scaling relations borrowed
from Cold Dark Matter
(``forwards'') models qualifies our approach rather as a ``hybrid'' one.
In fact, the scaling relations allow to tie the geometric properties of disks
to those of the Milky Way and nothing more; 
the overall disk evolution is really determined
by the assumed infall (or mass accretion) history.

\section{Results}
\label{sec:results}

Our ``hybrid'' approach allows to reproduce successfully a large number
of observables of galaxies in the local Universe (BP2000), including:
disk sizes and central surface brightness, Tully-Fisher relations in
various wavelength bands, colour-colour and colour-magnitude relations, gas
fractions vs magnitude, abundances as a function of local and integrated
properties etc; it also reproduces naturally detailed spectra from the
Kennicut atlas of galaxies for various galaxy types.

We find that the crucial ingredient of that success is the assumption that
gas is infalling more slowly in galaxies with smaller masses and/or surface 
densities. These two parameters are the main drivers of disk evolution.
We note that Bell and de Jong (2000) reach similar conclusions, but they
attribute a more important role to local surface density while we find that 
galaxy mass plays an even more important role: on average, massive disks are
older than lower mass ones. Disks with $V_C\sim$100 km/s have formed the
first half of their stars within 4-9 Gyr, while disks with $V_C\sim$300 km/s 
have done so within 2-6 Gyr (depending on spin parameter, that is 
surface density, see Boissier et al. 2001).

The main observational arguments on which our conclusion is based are presented
in Fig. 3. One sees that, on average, massive disks are redder, poorer in
gas and richer in metals than less massive ones.
Each one of these observables, taken separately, could be easily explained
(e.g. redder colours could be explained by larger amounts of dust, as invoked
e.g. in Somerville and Primack 1999 or Avila-Reese and Firmani 2000).
When all
three observables are considered together, the idea of a star formation
efficiency increasing with galaxy mass apears as a viable explanation.
Such a possibility has been invoked in Ferreras and Silk (2001), but it
requires huge amounts of (undetectable) gas to be present in small galaxies. 
However, when the fourth observable is taken into account, namely the
absence of any dependence of the present-day of star formation efficiency on 
rotational velocity (or B-magnitude, see Boissier et al. 2001), the only
viable explanation is the one invoked above: massive disks are older.

\section{Conclusion}
\label{sec:conclusion}

In the currently popular paradigm of hierarchical galaxy formation, low mass 
dark matter haloes form first, while more massive ones are formed later 
through accretion and merging; in principle, baryons are supposed to follow
the dark matter, but their fate is largely unknown at present, due to a lack 
of a reliable theory of star formation (and feedback). 

In recent cosmological
hydro-simulations by Nagamine et al. (2001) it is found that star formation in
small galaxies has stopped many Gyr ago, in clear contradiction with
observations of local galaxies. On the other hand, van den Bosch (2002) finds
that semi-analytical models, even with  feedback, produce massive
disks that are systematically bluer than their lower mass counterparts, again
in contradiction with observations; the reason of the failure is obviously
related to the fact that the mass accretion histories of baryons are largely
dictated by the hierarchical clustering of dark matter (e.g. Avila-Reese and Firmani
2000). Once
gas becomes available  it forms rapidly stars; feedback can only delay
star formation for a short time (shorter than the several Gyr that
are observationally required to obtain small disks bluer than massive ones).

Our simple, ``hybrid'' model for disk evolution, calibrated on the MW, 
suggests that on average massive disks have formed the bulk of their stars
several Gyr earlier than low mass ones. Their predictions match  
successfully most currently available observables, including data
from surveys at intermediate redshifts (Boissier and Prantzos 2001).

At present, and despite claims to the contrary, there is no satisfactory 
explanation (at least, not a published one) 
for the observables presented in Fig. 3 in the framework of
hierarchical galaxy formation. It remains to be shown why star formation
in galaxies apparently followed an ``inverted hierarchy'' w.r.t the dark
matter asembly. Feedback offers  an obvious solution to that problem, but
the required delay timescales appear unphysically large.

\end{document}